\documentclass[preprint,cite]{epl}
\usepackage{epsfig}
\begin{document}

\title{Wetting at Curved Substrates: Non-Analytic Behavior of Interfacial
Properties}
\shorttitle{Wetting at Curved Substrates}
\shortauthor{R. Evans \etal}
\author{R. Evans\inst{1,2,3} \and R. Roth\inst{2,3} \and  P. Bryk\inst{2,3,4}}
\institute{\inst{1}H.H. Wills Physics Laboratory, University of Bristol, 
Bristol BS8 1TL, UK\\
\inst{2}Max-Planck-Institut f{\"u}r Metallforschung, Heisenbergstr. 3,
70569 Stuttgart, Germany\\
\inst{3}ITAP, Universit{\"a}t Stuttgart, Pfaffenwaldring 57, 70569 
Stuttgart, Germany\\
\inst{4}Dep. of Modeling of Physico-Chemical Processes, MCS
University, Lublin 20-031, Poland}


\maketitle
\begin{abstract}
We argue that for complete wetting at a curved substrate (wall) the wall-fluid
surface tension is non-analytic in $R_i^{-1}$, the curvature of the wall and 
that the density profile of the fluid near the wall acquires a contribution
proportional to the gas-liquid surface tension $\times R_i^{-1}$ plus 
higher-order contributions which are non-analytic in $R_i^{-1}$. These 
predictions are confirmed by results of density functional calculations for
the square-well model of a liquid adsorbed on a hard sphere and on a hard
cylinder where complete wetting by gas (drying) occurs. The implications of our
results for the solvation of big solvophobic particles are discussed.
\end{abstract}

Understanding the adsorption of fluids at solid substrates has taken on new
importance with recent advances in the controlled fabrication of tailored
surfaces for applications in microfluidics and other areas \cite{Dietrich99}. 
Much experimental \cite{Bruschi02} and theoretical effort \cite{Rascon00} is 
concerned with wetting and associated interfacial transitions in wedge 
geometry and recently attention has turned to wetting at an apex 
\cite{Parry03}. It is becoming increasingly clear that substrate geometry can 
have a profound influence on the nature of fluid adsorption and, in particular,
on wetting characteristics making these quite different from those at a planar
substrate. Here we consider complete wetting at two substrates that have simple
geometries, namely a single sphere of radius $R_s$ and an infinitely long
cylinder of radius $R_c$. Using an effective interfacial Hamiltonian approach
\cite{Dietrich88} combined with exact microscopic sum-rules for the density 
profile of the fluid near a hard wall, we show that in the limit 
$R_i\to\infty$, with $i=s,c$, the surface tension of the substrate (wall)-fluid
interface is non-analytic in the curvature $R_i^{-1}$ and that the density of 
the fluid in contact with the hard wall acquires a contribution proportional to
$\gamma_{gl}(\infty)/R_i$, where $\gamma_{gl}(\infty)$ is the surface tension 
of the planar interface between coexisting gas and liquid, as well as 
higher-order non-analytic terms in $R_i^{-1}$. Our results, which are 
confirmed fully by the results of microscopic density functional (DFT) 
\cite{Evans92} calculations, show that non-zero curvature leads to unexpected 
and subtle effects on interfacial properties when complete wetting occurs, 
even for the simplest of substrate geometries.

In previous theoretical studies of wetting on spheres and cylinders 
\cite{Holyst87,Gelfand87,Upton89,Bieker98} the thrust was on understanding how
a finite radius limits the thickness of a wetting film and modifies the wetting
transition that can occur at a planar substrate. Little attention was paid to 
the effect of curvature on the surface tension and on the form of the density 
profile near the substrate which are the main concerns of this Letter. Our
results have repercussions for the general theory of solvation of big 
solvophobic solute particles, for wetting of colloidal particles and fibers
\cite{Dietrich99,Bieker98} and, possibly, for the surface tension of drops and 
bubbles \cite{Rowlinson94}.

We consider first the general case of a bulk fluid phase $a$, at chemical
potential $\mu$, in contact with a wall $w$ and implement a standard, 
coarse-grained, effective interfacial Hamiltonian approach 
\cite{Dietrich88,Gelfand87} in which the wetting film of fluid phase $b$ is 
characterized only by its thickness $l$. For short-ranged (finite ranged, 
exponentially or faster decaying) wall-fluid and fluid-fluid potentials the
binding potential, i.e. the excess (over bulk) grand potential per unit area, 
takes the form 
\cite{Dietrich88,Gelfand87}:
\begin{equation} \label{curved}
\tilde\omega_{wa}^i(l)=\gamma_{wb}^i(R_i)+\gamma_{ba}^i(R_i+l) + 
\tilde a(T) e^{-l/\xi} + (\rho_a-\rho_b) \delta \mu~ l +
\frac{\alpha_i \gamma_{ba}^i(R_i+l) l}{R_i} + {\cal O}(l/R_i)^2,
\end{equation}
with $i=s,c$ and $\alpha_s=2$, $\alpha_c=1$ for the spheres and cylinders, 
respectively. $\gamma_{wb}^i$ is the surface tension of the wall-phase $b$
interface, $\gamma_{ba}^i$ is the tension of the $ba$ fluid-fluid interface
located near $R_i+l$, $\tilde a(T)>0$ is a coefficient that we need not specify
further for complete wetting, and $\xi$ is the true bulk correlation length of
the wetting phase $b$. The fourth term in eq.~(\ref{curved}) accounts for the
increase in grand potential per unit area associated with the volume of a film
of phase $b$; $\delta \mu \equiv \mu-\mu_{co}(T)$ is the chemical potential
deviation and $\rho_a$, $\rho_b$ are the number densities of bulk phases $a$
and $b$ at coexistence $\mu_{co}(T)$. For $R_i=\infty$ the binding potential 
(\ref{curved}) reduces to that appropriate for complete wetting at a planar
interface in models where the range of the wall-fluid potential is shorter 
than $\xi$ and, henceforward, we shall assume this is the case. Then minimizing
eq.~(\ref{curved}) w.r.t. $l$ yields the well-known logarithmic divergence of
the equilibrium film thickness:
$l_{eq}(\infty)=-\xi\ln\left(\delta \mu~\xi (\rho_a-\rho_b)/\tilde a(T)\right)$,
as $\delta \mu \to 0$.
When $R_i\not= \infty$ two important modifications arise: i) $\gamma_{wb}^i$
and $\gamma_{ba}^i$ now depend on curvature and ii) the surface area of the
$ba$ interface now depends on the film thickness $l$. Equation~(\ref{curved})
assumes very large radii $R_i$ and that $l/R_i\ll 1$. Note that the fifth term
in (\ref{curved}) is 
proportional to the Laplace pressure across the fluid-fluid $ba$ interface. 
Since $\gamma_{ba}$ is always non-zero, away from the bulk critical point, 
this term ensures that the film thickness remains finite even at $\mu_{co}(T)$,
i.e. minimizing eq.~(\ref{curved}) yields 
$l_{eq}^i(R_i)=-\xi \ln[\alpha_i \xi \gamma_{ba}^i(R_i)/(R_i \tilde a(T))]$
for $\delta \mu =0$, where we have ignored terms ${\cal O}(l/R_i)^2$ and the
$l$ dependence in $\gamma_{ba}^i$ -- this can be justified a posteriori. 
Several authors \cite{Holyst87,Gelfand87,Upton89,Bieker98} have argued that 
for large $R_i$ the quantity $\alpha_i \gamma_{ba}^i(R_i)/R_i$ should play the
same role in complete wetting at a curved wall, with $\delta \mu=0$, as the 
effective bulk field $(\rho_a-\rho_b) \delta \mu$ plays for the planar wall. 
We shall pursue this argument further in the present Letter.

The wall-fluid surface tension is given by
\begin{equation}
\gamma_{wa}^i(R_i)\equiv \tilde \omega_{wa}^i(l_{eq}) = \gamma_{wb}^i(R_i)+
\gamma_{ba}^i(R_i)+(\xi+l_{eq})\left(\frac{\alpha_i \gamma_{ba}^i(R_i)}{R_i}
+(\rho_a-\rho_b) \delta \mu \right).
\end{equation}
For the planar interface the final term vanishes at $\delta \mu =0$ and the
wall-fluid tension reduces to the sum of the planar tensions:
$\gamma_{wa}(\infty)=\gamma_{wb}(\infty)+\gamma_{ba}(\infty)$, appropriate to
complete wetting by a macroscopic film of phase $b$. When $\delta \mu \not= 0$,
$\gamma_{wa}(\infty)$ acquires a non-analytic contribution proportional to
$-|\delta \mu| \ln |\delta \mu|$ \cite{Dietrich88}. For finite $R_i$ we set 
$\delta \mu = 0$ and obtain
\begin{equation} \label{gammacurved}
\gamma_{wa}^i(R_i)=\gamma_{wb}^{i}(R_i)+\gamma_{ba}^i(R_i)\left(1+
\frac{\alpha_i \xi}{R_i}\right)+\frac{\alpha_i \xi \gamma_{ba}^i(R_i)}{R_i}
\ln(R_i \times const) +H.O.T..
\end{equation}
That the surface tension contains a contribution which is non-analytic in the
curvature is clearly a direct manifestation of complete wetting. However, the
necessity for such a contribution does not seem to have been widely recognized.
An exception is Ref.~\cite{Holyst87} but in that paper the consequences were 
not discussed. In order to elucidate these we must examine the other terms in 
eq.~(\ref{gammacurved}). At the non-wet $wb$ interface we do not expect 
non-analyticities in $\gamma_{wb}^i(R_i)$, provided the system exhibits 
short-ranged forces \cite{Bryk03}. For a spherical fluid-fluid interface it is
usually assumed \cite{Rowlinson94} (for short-ranged forces) that
\begin{equation} \label{hendersons}
\gamma_{ba}^s(R_s)=\gamma_{ba}(\infty) (1-2 \delta_T^s/R_s+H.O.T.),
\end{equation}
where $\delta_T^s$ is the Tolman length \cite{Tolman49} familiar in studies of
liquid drops, whereas for cylindrical interfaces it has been argued 
\cite{Henderson84} that
\begin{equation} \label{hendersonc}
\gamma_{ba}^c(R_c)=\gamma_{ba}(\infty)(1+b_H \ln R_c/R_c + H.O.T.).
\end{equation}
If the latter form were correct eq.~(\ref{gammacurved}) would imply
that the wall-fluid tension $\gamma_{wa}^{c}(R_{c})$ should have
a contribution $\gamma_{ba}(\infty)(b_{H}+\xi)\ln(R_{c})/R_{c}$.
Provided the length $b_{H}$ is comparable with the bulk correlation length 
$\xi$, as is expected on physical grounds, the term in $b_{H}$ should be 
easily identified in numerical work. 

The result (\ref{gammacurved}) is valid for any complete wetting situation, 
with the proviso that the forces are short-ranged. We specialize now to the 
case of a {\em hard} wall exerting a purely repulsive potential on the fluid: 
$V_{i}(r)=\infty$ for radius $r<R_{i}$ and is zero for $r>R_{i}$. It is 
well-known that the phenomenon of complete {\em drying} occurs at a planar hard
wall: the interface between the bulk liquid ($l$) and the wall is wet by a 
macroscopic film of gas ($g$) as $\mu\to\mu_{co}^{+}(T)$ for all temperatures 
$T$ at which gas and liquid coexist \cite{Henderson85,Tarazona84}. At  
$\mu=\mu_{co}^{+}(T)$ the density profile of the fluid is a composite of the 
planar wall-gas and the (free) gas-liquid interfacial profiles and 
$\gamma_{wl}(\infty)=\gamma_{wg}(\infty)+\gamma_{gl}(\infty)$; phase 
$a\equiv l$ and phase $b\equiv g$. Whilst drying at a planar hard wall has been
investigated extensively using computer simulations \cite{Henderson85} and DFT 
\cite{Tarazona84}, we are not aware of studies on curved substrates. We shall 
see below that curvature leads to some striking new results. When 
$R_{i}\neq\infty$ thick films of gas will still develop at the wall but the 
thickness will remain finite at $\mu_{co}^{+}(T)$ and the wall-liquid tension 
$\gamma_{wl}^{i}(R_{i})$ should exhibit the non-analyticities described above. 
Moreover the density profile $\rho(r)$ is no longer a perfect composite of the
two separate interfacial profiles and exhibits interesting features near the 
wall. 

A particular advantage of hard walls is that there is an exact statistical 
mechanical sum-rule \cite{Henderson86} which relates the fluid density at 
contact, $\rho^{i}(R_{i}^{+})$, to the pressure  $p$ of the bulk (reservoir) 
fluid and the wall-fluid tension:
\begin{equation}\label{sumrule1}
k_{B}T\rho^{i}(R_{i}^{+})=p+\frac{\alpha_i\gamma_{wf}^i(R_i)}{R_i}+
\left(\frac{\partial\gamma_{wf}^i(R_i)}{\partial R_i}\right)_{T,\mu}.
\end{equation}
In the limit $R_i=\infty$ the contact density reduces to $p/k_{B}T$, the 
well-known planar contact theorem. Equation~(\ref{sumrule1}) is valid for 
{\em any} one-component fluid at a hard wall. We now insert our result 
(\ref{gammacurved}) for the tension of the `dry' interface, 
$\gamma_{wl}^{i}(R_{i})$, into (\ref{sumrule1}) to obtain the contact density 
$\rho_{liq}^{i}(R_{i}^{+})$ at $\mu=\mu_{co}^{+}(T)$. By subtracting 
$\rho_{gas}^{i}(R_{i}^{+})$, the contact density when the bulk phase is gas at
$\mu=\mu_{co}^{-}(T)$, we eliminate both the pressure $p(\mu_{co})$ and the 
(analytic) wall-gas tension obtaining for a hard spherical wall:
\begin{equation}\label{sumrule2}
k_{B}T(\rho_{liq}^{s}(R_{s}^{+})-\rho_{gas}^{s}(R_{s}^{+}))=
\frac{2\gamma_{gl}(\infty)}{R_{s}}+\frac{2\xi\gamma_{gl}(\infty)}
{R_{s}^{2}}\ln R_{s}+H.O.T.
\end{equation}
and for a hard cylindrical wall:
\begin{equation}\label{sumrule3}
k_{B}T(\rho_{liq}^{c}(R_{c}^{+})-\rho_{gas}^{c}(R_{c}^{+}))=
\frac{\gamma_{gl}(\infty)}{R_{c}}+\frac{(\xi+b_{H})\gamma_{gl}(\infty)}
{R_{c}^{2}}+H.O.T.
\end{equation}
where $H.O.T.$ refers to terms of higher order in $R_i^{-1}$. For a planar wall
the contact density is $p(\mu_{co})/k_{B}T$ in {\em both} phases so the 
difference vanishes identically. There are two remarkable features in these 
results. First the contact density in the liquid phase (`dry' interface) 
depends on $\gamma_{gl}(\infty)$, the tension of the planar gas-liquid 
interface, which can be very far from the wall as $R_{i}\to\infty$: the first 
term on the r.h.s. of (\ref{sumrule2}) and (\ref{sumrule3}) is simply the 
Laplace pressure across the gas-liquid interface. Second the contact density 
difference for the sphere contains a next to leading order 
$R^{-2}_{s}\ln R_{s}$ non-analyticity, whereas for the cylinder the next to 
leading order term is analytic, i.e. ${\cal O}(R^{-2}_{c})$, despite the fact 
that both surface tensions are non-analytic at order $R^{-1}_{i}\ln R_{i}$.

\begin{figure}
\onefigure[width=6.5cm]{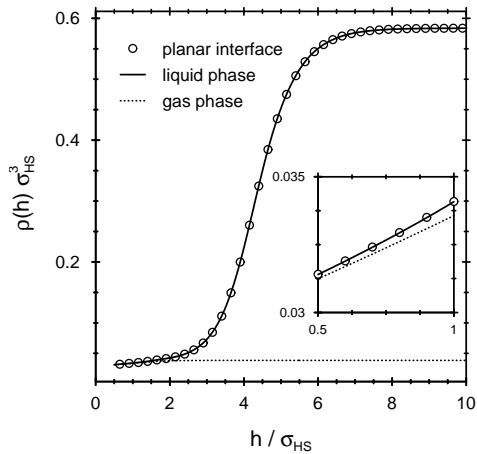}

\caption{\label{fig:profile} The density profile of a square-well fluid 
adsorbed at a hard wall for $k_B T/\varepsilon=1$. Solid curve refers to a
liquid at bulk coexistence $\mu_{co}^+(T)$ near a spherical wall of radius
$R_s=2500\sigma_{HS}$ and the dotted curve to the gas at $\mu_{co}^-(T)$ at the
same wall. Symbols refer to the liquid near a planar wall at chemical potential
$\mu-\mu_{co}(T)=2 \gamma_{gl}(\infty)/(R_s(\rho_l-\rho_g))$ (see text). 
One observes that the two wall-liquid profiles are indistinguishable in the
region of the gas-liquid interface {\em and} close to the wall -- see inset. 
Contact occurs at distance $h=\sigma_{HS}/2$.}
\end{figure}

In order to test the predictions of the coarse-grained theory we have adopted 
a fully microscopic DFT approach and performed calculations for a square-well 
fluid at hard curved walls. The fluid-fluid pair potential $\phi (r)$ is 
infinite for $r<\sigma_{HS}$, the hard sphere diameter, the width of the well 
is 0.5$\sigma_{HS}$ and the well-depth is $\varepsilon$. We treat the 
hard-sphere part of the free-energy functional by means of Rosenfeld's 
fundamental measure theory \cite{Rosenfeld89} and the attractive part using a 
simple mean-field approximation \cite{Evans92,Tarazona84}, taking 
$\phi_{att}(r)=-\varepsilon$ for $r<1.5 \sigma_{HS}$ and zero otherwise. By 
minimizing the grand potential functional we determine the equilibrium density
profile $\rho(r)$ and the wall-fluid tension for any thermodynamic state;
details of the numerical treatment will be given elsewhere. Our DFT approach
has the following advantageous features: i) the coexisting densities of gas and
liquid, $\rho_g$ and $\rho_l$, can be calculated precisely, ii) the DFT obeys
the sum-rule (\ref{sumrule1}) and the Gibbs adsorption theorem and iii) the
key quantities $\xi$ and $\gamma_{gl}(\infty$) which enter 
eqs.~(\ref{gammacurved},\ref{sumrule2},\ref{sumrule3}) can be obtained from
{\em independent} calculations (for the bulk and for the planar interface) 
using the same functional.

We focus first on the density profile. In fig.~\ref{fig:profile} we display
results for $\rho(r)$, at a reduced temperature $k_B T/\varepsilon = 1$ and
chemical potential $\mu_{co}^+(T)$, for the square-well fluid adsorbed on a 
hard sphere with radius $R_s=2500 \sigma_{HS}$. In the same figure we show the
profiles for the wall-gas interface at $\mu^-_{co}(T)$ and for the planar 
interface with 
$\delta\mu\equiv\mu-\mu_{co}(T)=2\gamma_{gl}(\infty)/(R_s (\rho_l-\rho_g))$.
One can observe that this translation between bulk field for a planar wall and
curvature yields wall-liquid profiles which lie on top of each other. This is a
non-trivial result. Although it is evident that within the coarse-grained 
treatment of eq.~(\ref{curved}) the film thickness $l_{eq}^i$ is identical for
planar and curved walls, there is no reason a priori why the complete 
microscopic density profiles should be especially close. 

\begin{figure}
\twofigures[width=6.5cm]{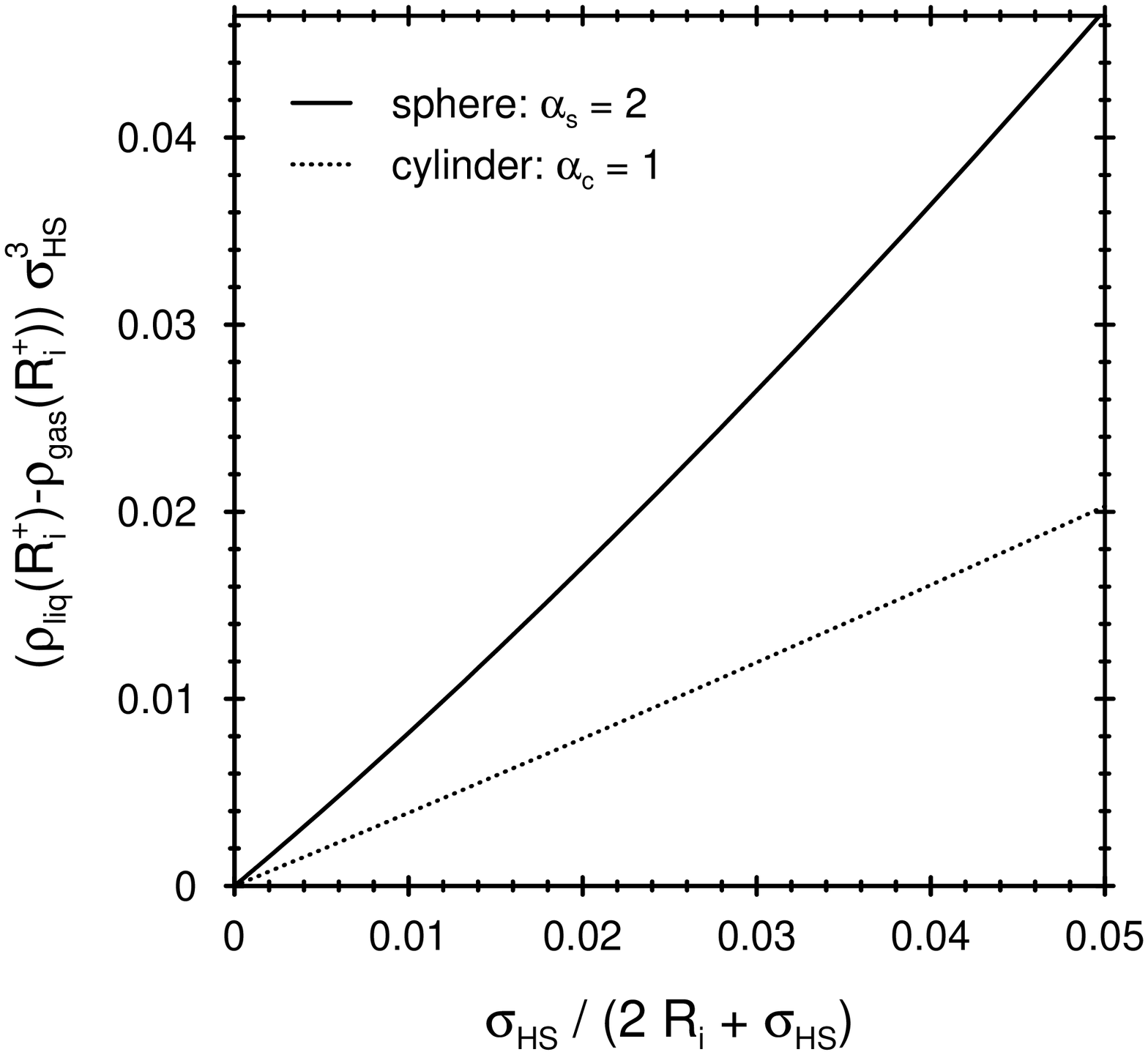}{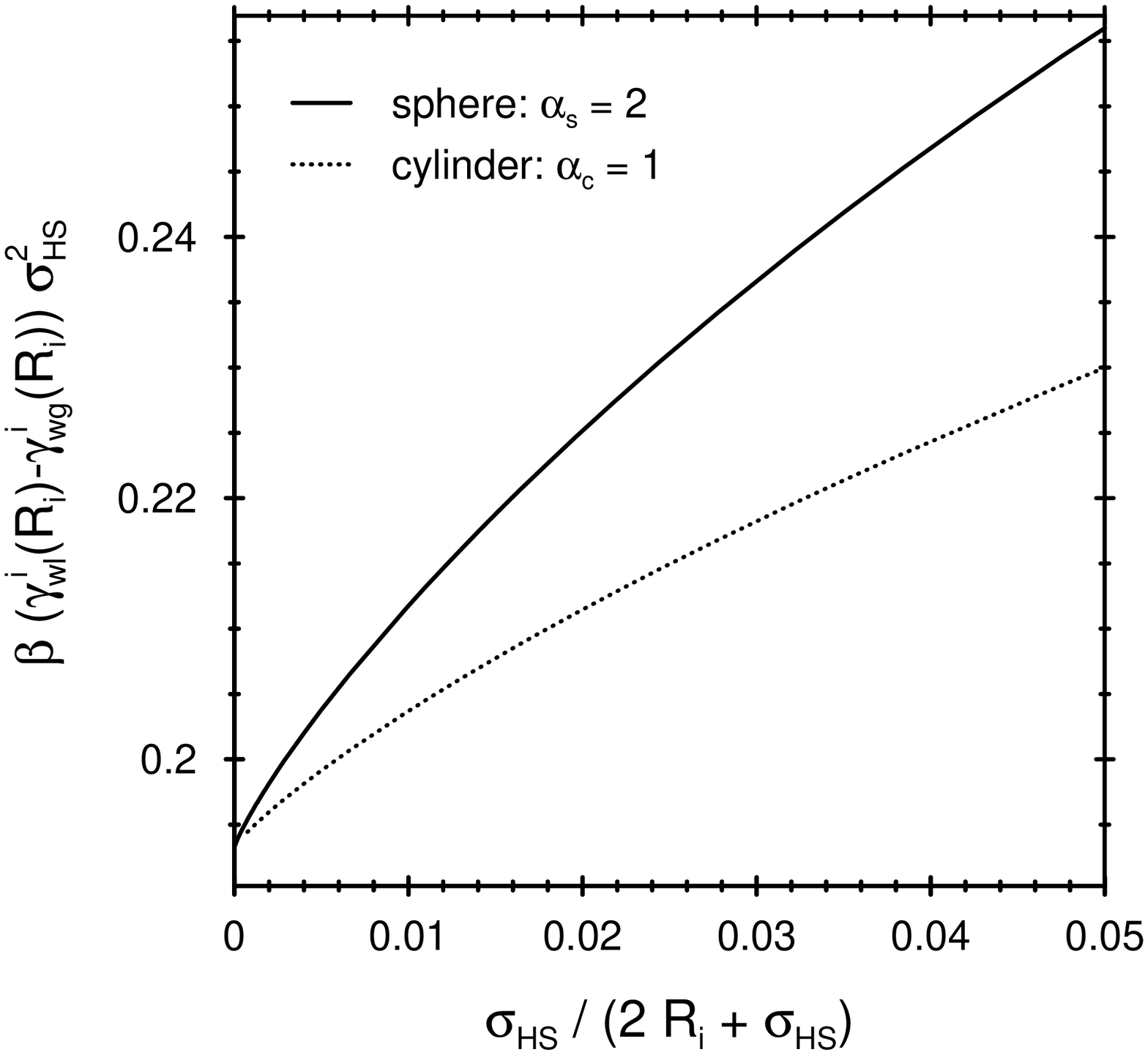}

\caption{\label{fig:contact} The difference between contact densities of a 
square-well fluid in the liquid and in the gas phases, at $\mu_{co}^{\pm}(T)$,
adsorbed at a hard spherical (full line) and hard cylindrical (dotted line) 
wall of radius $R_i$ for $k_B T/\varepsilon=1$. From the linear portion of each
curve near the origin, we can extract the gas-liquid surface tension 
$\gamma_{gl}(\infty)$. The next to leading order term is proportional to
$R_s^{-2} \ln R_s$ for the sphere and to $R_c^{-2}$ for the cylinder --
see eqs.~(\protect\ref{sumrule2}) and (\protect\ref{sumrule3}).}

\caption{\label{fig:gamma} As in fig.~\protect\ref{fig:contact} but now for the
difference between the surface tensions. For both geometries the leading order
curvature dependence is proportional to $\ln R_i/R_i$ -- see
eq.~(\protect\ref{gammacurved}). In both cases the coefficient is confirmed to
be $\alpha_i \xi \gamma_{gl}(\infty)/k_B T$, with $\xi$ the correlation length
in the bulk gas.}
\end{figure}

In fig.~\ref{fig:contact} we seek to test the validity of eqs.~(\ref{sumrule2},
\ref{sumrule3}) by examining the difference in the contact densities for the
two phases as a function of $R_i^{-1}$. Our DFT results are in excellent
agreement with the prediction of the coarse grained theory for both spheres
and cylinders. In particular we find the data is fit best with a 
$R_s^{-2} \ln R_s$ contribution for spheres and without such a contribution
for cylinders. It is straightforward to extract the coefficient of the
$R_i^{-1}$ contribution and we find close agreement, to 1 part in $10^4$, with
our independent (planar) results for the dimensionless quantity 
$\alpha_i \gamma_{gl}(\infty) \sigma_{HS}^2/k_B T$.

In fig.~\ref{fig:gamma} we plot the difference between the wall-liquid and
wall-gas surface tensions, evaluated at $\mu_{co}^\pm(T)$, respectively versus
$R_i^{-1}$ for the square-well fluid adsorbed at spheres and cylinders, again
for $k_B T/\varepsilon = 1$. In accordance with eqs.~(\ref{gammacurved}) and
(\ref{hendersons}) we expect that for the sphere the leading order curvature
correction to the difference is proportional to $R_s^{-1} \ln R_s$ with a
coefficient given by $2 \xi \gamma_{gl}(\infty)$ where $\xi$ is the 
correlation length in the bulk gas. For cylinders we employ
eqs.~(\ref{gammacurved}) and (\ref{hendersonc}) which imply the same type of
leading order correction but now with a coefficient 
$(\xi+b_H)\gamma_{gl}(\infty)$. Our numerical DFT results are completely
consistent with these predictions with the relevant coefficients agreeing with the planar
result to better than $1 \%$. In our calculations we were able to obtain 
reliable results for $R_s$ up to $30000\sigma_{HS}$ and $R_c$ up to 
$5000\sigma_{HS}$. For the cylinder our best fit yields a coefficient that has 
$b_H = 0$ (or a very small fraction of $\xi$) and we conclude that there is no 
evidence for the $\ln R_c/R_c$ term conjectured \cite{Henderson84} for the 
cylindrical fluid-fluid interface -- see eq.~(\ref{hendersonc}). We have 
confirmed that the same level of agreement between the coarse-grained theory 
and DFT holds at other temperatures.

Since there is nothing special about the square-well model our predictions for
hard spheres and cylinders should be valid for any fluid with short-ranged
fluid-fluid potentials that exhibits gas-liquid coexistence and this has
important repercussions for the solvation of big solvophobic solute particles.
The excess chemical potential of a single hard sphere of radius $R_s$ in a
solvent is given generally by 
$\tilde\mu^{HS}(R_s)=p 4\pi R_s^3/3+\gamma_{wf}^s(R_s) 4 \pi R_s^2$, where
$\gamma_{wf}^s$ is the wall-fluid surface tension. Using 
eq.~(\ref{gammacurved}) it follows that the difference between 
$\tilde\mu^{HS}$ evaluated in the liquid, where drying occurs, and in the gas
phase, at $\mu_{co}^\pm(T)$, is
\begin{equation}
\frac{(\tilde\mu_{liq}^{HS}-\tilde\mu_{gas}^{HS})}{4 \pi R_s^2} =
\gamma_{gl}(\infty)(1+2 \xi R_s^{-1} \ln R_s) + H.O.T.,
\end{equation}
for $R_s\to\infty$. Whilst it is known from recent studies \cite{Lum99} of 
hard-sphere solutes that $\tilde\mu_{liq}^{HS}$ should contain a gas-liquid 
surface tension contribution $\gamma_{gl}(\infty) 4 \pi R_s^2$, the presence 
of the non-analytic correction is striking -- especially when we recall that 
the excess chemical potential is the derivative of the excess free energy of 
the {\em bulk mixture} w.r.t. the solute density in the limit of vanishing 
solute.

The coarse grained treatment that leads to eq.~(\ref{gammacurved}) is valid
for wetting by either fluid phase. Thus, provided the attractive part of the
wall-fluid potential is sufficiently strong and has finite range or decays
on a length scale that is shorter than the bulk correlation length $\xi$, now
referring to the wetting liquid, adsorption from the saturated gas, at
$\mu_{co}^-(T)$, will give rise to a wall-gas tension which has a term
$\alpha_i \xi \gamma_{gl}(\infty) R_i^{-1} \ln R_i$. Although the contact 
theorem (\ref{sumrule1}) is modified when the wall-fluid potential is no 
longer purely hard the contact density in the presence of a wetting liquid 
film will still acquire terms which depend on the gas-liquid surface tension, 
i.e. we expect results equivalent to (\ref{sumrule2}) and (\ref{sumrule3}).

Both the coarse-grained and DFT approaches presented here are mean-field like 
in that they omit effects of capillary-wave fluctuations in the wetting film
\cite{Dietrich88,Evans92}. For complete wetting in three dimensions at a planar
interface, renormalization group studies based on effective interfacial 
Hamiltonians with the binding potential (\ref{curved}) find that critical 
exponents are not altered from their mean-field values but the amplitude of the
equilibrium film thickness $l_{eq}(\infty)$ is changed from $\xi$ to 
$\xi(1+\omega/2)$, for $\omega<2$, when fluctuations are included 
\cite{Dietrich88}. Here $\omega=k_B T/(4 \pi \gamma_{gl}(\infty)\xi^2)$ is the
usual parameter which measures the strength of capillary-wave fluctuations: 
$\omega=0$ corresponds to mean-field. We conjecture that our present results 
for the leading non-analytic term in the surface tension are modified in a 
similar fashion, i.e. the third term in eq.~(\ref{gammacurved}) simply has 
$\xi$ replaced by $\xi (1+\omega/2)$. As regards the results (\ref{sumrule2}) 
and (\ref{sumrule3}) for the difference in the contact densities, the 
leading-order (Laplace pressure) terms will be unchanged by fluctuations 
whereas the next to leading order term in eq.~(\ref{sumrule2}) will require 
the same replacement for $\xi$.

For real fluids dispersion forces are always present giving rise to $r^{-6}$
power-law decay of the fluid-fluid pair potential. This leads, in turn, to a
wetting film thickness which diverges at coexistence as 
$l_{eq}(R_i)\sim R_i^{1/3}$, for $R_i\to\infty$ \cite{Bieker98,Brochard86}.
The present coarse-grained analysis suggests that the non-analyticities in 
curvature which arise for wetting with such potentials will be power laws 
rather than terms involving logarithms and we are presently investigating 
these using DFT \cite{comment}.

We have shown that the isomorphism between bulk field 
$\delta \mu(\rho_l-\rho_g)$ for a planar substrate and Laplace pressure at a
curved substrate implied by the binding potential (\ref{curved}) leads to 
striking consequences for interfacial properties: first one cannot obtain the
surface excess free energy (surface tension) of a fluid that wets completely a
non-planar substrate by expanding only in powers of the curvature(s). Second
the true microscopic density profile near the curved substrate depends on the
surface tension of the gas-liquid interface which, for large $R_i$, can be
far from the substrate. A detailed explanation of this curious behavior will
be given elsewhere. Here we merely state that it is associated with the 
exponential decay (for short-ranged potentials) of the tails of the density
profile of the `free' gas-liquid interface. This is not the first time that
analysis of complete wetting or drying at a hard wall has caused surprises and
lead to new insight into the fundamental physics of fluid interfaces
\cite{Henderson85,Parry93}.

We have benefited from conversations with J.R. Henderson and M. Thomas. R.E.
is grateful to S. Dietrich for kind hospitality and to the Humboldt Foundation
for support under GRO/1072637 during his stay in Stuttgart.


\begin{thebibliography}{0}
\bibitem{comment} Since the upper critical dimension for complete wetting with
power law potentials is $<3$, fluctuations are not important in this case.
\end{thebibliography}
\end{document}